\documentclass[aps,prb,twocolumn,showpacs,superscriptaddress,groupedaddress]{revtex4}  % for review and submission
\usepackage{graphicx}  % needed for figures
\usepackage{dcolumn}   % needed for some tables
\usepackage{bm}        % for math
\usepackage{amssymb}   % for math
\usepackage{changes}

\usepackage{amsmath}    % need for subequations
\usepackage{color}      % use if color is used in text
\usepackage{subfigure}  % use for side-by-side figures

\usepackage{ulem}  % needed for non standard text attributes

\begin{document}

% The following information is for internal review, please remove them for submission
\widetext
%\leftline{Version xx as of \today}
%\leftline{Primary authors: Joe E. Physics}
%\leftline{To be submitted to (PRL)}
%\leftline{Comment to {\tt d0-run2eb-nnn@fnal.gov} by xxx, yyy}
%\centerline{\em D\O\ INTERNAL DOCUMENT -- NOT FOR PUBLIC DISTRIBUTION}

% the following line is for submission, including submission to the arXiv!!
%\hspace{5.2in} \mbox{Fermilab-Pub-04/xxx-E}

%\title{Scalable interconnections for remote exciton systems}
\title{Coherent transport and manipulation of spins in indirect exciton nanostructures}
%\input author_list.tex       % D0 authors (remove the first 3 lines
                             % of this file prior to submission, they
                             % contain a time stamp for the authorlist)
                             % (includes institutions and visitors)


\author{A. Violante}
\email[e-mail address: ]{violante@pdi-berlin.de}
\affiliation{Paul-Drude-Institut f\"ur Festk\"orkperelektronik, Hausvogteiplatz 5-7, 10117 Berlin, Germany}
\author{R. Hey}
\affiliation{Paul-Drude-Institut f\"ur Festk\"orkperelektronik, Hausvogteiplatz 5-7, 10117 Berlin, Germany}
\author{P. V. Santos}
\affiliation{Paul-Drude-Institut f\"ur Festk\"orkperelektronik, Hausvogteiplatz 5-7, 10117 Berlin, Germany}
%\thanks{}

\date{\today}

\begin{abstract}
 We report on the coherent control and transport of indirect exciton (IX) spins in GaAs double quantum well (DQW) nanostructures. The spin dynamics was investigated by optically generating spins using a focused, circularly polarized light spot and by probing their spatial distribution using spatially and polarization resolved photoluminescence spectroscopy. Optically injected exciton spins precess while diffusing over distances exceeding  $20~\mu$m from the excitation spot with a spatial precession frequency  that depends on the spin transport direction as well as on the bias applied  across the DQW structure. This behavior is attributed to the spin precession in the effective magnetic field induced by the spin-orbit interaction. From the dependence of the spin dynamics on the transport direction, bias and  external magnetic fields we directly determined the  Dresselhaus and Rashba spin splitting coefficients for the structure. The precession dynamics is essentially independent on the IX density, thus indicating that the long spin lifetimes are not associated with IX collective effects. The latter, together with the  negligible contribution of holes to the spin dynamics, are rather attributed to  spatial separation of the electron and hole wave functions by the electric field, which reduces the electron-hole exchange interaction. Coherent spin precession over long transport distances as well as the  control of the spin vector using electric and magnetic field open the way for the application of \textit{IX} spins in the quantum information processing. 
\end{abstract}

\maketitle  %\maketitle must follow title, authors, abstract and \pacs

%\section{\label{sec:level1}First-level heading}
% sections are not used for PRL papers
%\label{}
%\subsection{}
%\subsection{\label{sec:level2}Second-level heading: Formatting}
% subsections are not used for PRL papers

\section{Introduction}
\label{Introduction}

\begin{figure}[tbhp]
\begin{center}
\includegraphics[width=\linewidth, keepaspectratio=true,angle=0,clip]{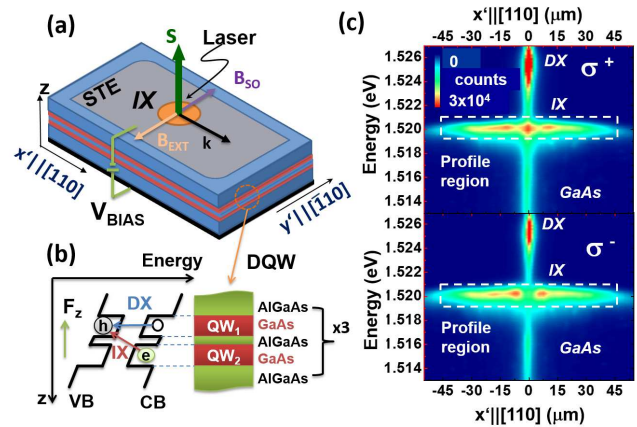}
\end{center}
\caption{(a) Double quantum well (DQW) structures  for indirect exciton (\textit{IX}) spin transport. The \textit{IX}s are optically excited using a focused laser pulse in a DQW subject to an electric field ($F_z$) induced by a bias $V_\mathrm{BIAS}$ applied between the doped substrate and a semitransparent top electrode (STE).  \textit{IX} spin transport is mapped by  imaging the IX photoluminescence (PL) emitted along direction $k$ with polarization sensitivity. During motion along $k$, the spins precess under the spin-orbit ($B_{SO}$) and external magnetic field ($B_{\mathrm{ext}}$). (b) Energy band diagram of the GaAs/(Al,Ga)As DQW along the vertical ($z$) direction showing the direct (\textit{DX}) and \textit{IX} transitions under the electric field $F_z$. 
(c) Spatially and spectrally resolved images of the PL with right ($\sigma^+$, upper panel) and left ($\sigma^-$, lower panel) circular polarizations recorded under a bias $V_\mathrm{BIAS}=-2$~V and a laser power of $13~\mu$W. \textit{DX} and \textit{IX} denote the emission from direct and indirect excitons in the DQW structure, while the line at 1.515~eV arises from excitons in the GaAs substrate.}
\label{FigIXSetup}
\end{figure}

An indirect (or dipolar) exciton (\textit{IX}) consists of a Coulomb-correlated electron-hole pair with the individual particles confined in  two closely spaced quantum wells (QWs) in a  double quantum well (DQW) structure (cf.~Fig.~\ref{FigIXSetup}) subjected to a transverse electric field $F_z$. $F_z$  controls via the quantum confined Stark effect both the \textit{IX} energy and radiative lifetime, which can reach the  ms-range.\cite{Sivalertporn_PRB85_45207_12}  
The electric field can be used to laterally confine \textit{IX} gases in  DQW plane using electrostatic gates~\cite{Schinner_PRL110_127403_13} as well as to control the interconversion between \textit{IX} and photons. In addition, the electrical dipoles arising from the charge separation give rise to  strong repulsive exciton-exciton interactions \cite{PVS239}. Finally, \textit{IX}s are composite bosons and, therefore, susceptible to the formation of collective exciton phases, as evidenced by the experiments presented in Refs.~\onlinecite{Butov_N418_751_02,High_Nature_12,Shilo_NC4_2335_13}.  
These interesting \textit{IX} properties have been combined to demonstrate different device functionalities including  light storage cells, switches, and transistors~\cite{High_S321_229_08,Winbow_NL7_1349_07,Grosso_NP3_577_09}. The long \textit{IX} lifetimes also enable the long-range transport and coupling  of  \textit{IX} packets. Different approaches have also been introduced for the transport of  \textit{IX}s  based on bare diffusion,\cite{Voros_PRL94_226401_05} drift induced by repulsive \textit{IX-IX} interactions,\cite{Butov_N418_751_02} or by using spatially varying electric fields, including electrostatic ramps~\cite{Gaertner_APL89_052108_06} as well as moving electrostatic~\cite{Winbow_PRL106_196806_11} and acoustic lattices~\cite{PVS177,PVS260,PVS266}.

Recently, the confinement of single \textit{IX}s has been demonstrated,~\cite{Schinner_PRL110_127403_13} thus opening the way for applications of single  \textit{IX}s in quantum information.~\cite{Zutic_ROMP76_323_04} 
Here, the manipulation and transport of \textit{IX} spins offers further exciting perspectives.  \textit{IX}s have several favorable properties for spin control and manipulation.  The reduced overlap between the electron and hole wave functions reduces not only the radiative lifetime but also the exchange interaction between the exciton-bound electron spins and the short-living hole spins (the excitonic or Bir-Aronov-Pikus (BAP) mechanism~\cite{Bir_SovPhysJETP42_705_75,Bir74a}), thus enhancing the exciton spin lifetime.~\cite{Larionov_JL71_117_00,Larionov_PRB73_235329_06,Kowalik-Seidl_APL97_11104_10} 

The long lifetime  also enables for spin control using the spin-orbit (SO) coupling. Due to the Bulk Inversion Asymmetry (BIA) intrinsic to the  noncentrosymmetric III-V materials (such as GaAs),\cite{Dyakonov_SPJETP_33_1053, Dyakonov_Springer-Verlag_08}  moving spins feel an effective magnetic field (the Dresselhaus field,\cite{Dresselhaus55a} $\bf B_D$), which modifies the spin vector during precession. In addition, the Structural Inversion Anisotropy (SIA) created by $F_z$ induces a tunable contribution to the SO magnetic field (the Rashba field,\cite{Rashba_SPSS2_1109_60} $\bf B_R$), thus providing a powerful tool for the dynamic spin control during motion. $F_z$ also affects spins by changing the $g$-factor.\cite{Poggio_PhysRevB70_121305_04} The SO fields can, however, also lead to the dephasing of exciton spin ensembles (the D'yakonov-Perel'  dephasing mechanism). The spin dephasing time can be electrically controlled by selecting  $F_z$ to modulate the total SO field $\bf B_{SO}=B_D+D_R$.\cite{Larionov_PRB78_33302_08} 
The \textit{IX} spin lifetime also enhances under high particle densities.   In analogy with the spin dynamics of atomic condensates,\cite{Burt_PRL79_337_97} the latter has been attributed to the formation of collective \textit{IX} phases.\cite{Timofeev_PhysRevB60_8897_99,Larionov_PRB73_235329_06,Leonard_NanoLett9_4204_09,High_PRL110_246403_13,High_Nature_12} 
In particular, experiments carried out in this regime have lead to the observation of spin currents in \textit{IX} gases with a variety of spatially resolved polarization patterns,\cite{High_PRL110_246403_13,Kavokin_PRB88_195309_13} which  have been attributed to the SO magnetic field.

Most of the previous \textit{IX} spin studies have addressed the spin dynamics very close to the spin excitation area  (i.e., within distances $<10~\mu$m), one exception being the self-organized extended (i.e., $>20~\mu$m) spin patterns reported for the collective regime at high \textit{IX} densities.\cite{High_Nature_12} 
%One exception are the self-organized extended (i.e., $>20~\mu$m), self-organized spin patterns reported for the collective regime at high \textit{IX} densities.\cite{High_Nature_12} 
%
In this work, we demonstrate that 
% report on the coherent transport of \textit{IX} spins in (001) GaAs DQW structures  investigated by spatially resolved, polarization sensitive photoluminescence (PL). O
optically injected \textit{IX} spins can be transported over distances exceeding 20~$\mu$m from the injection spot. The spins coherently precess during motion: the spatial precession period depends on the motion direction and can be controlled by the vertical field $F_z$ applied across the DQW. This precession behavior, which does not depend on \textit{IX} density, is attributed to the precession of the electron spins in the anisotropic SO field.  %The spatial dependence of the spin profiles is well reproduced taking into account the BIA and SIA SO contributions, from which we experimentally determined the Dresselhaus and Rashba spin-splitting coefficients. 
%The spin dynamics during diffusion can be further modified via an in-plane external magnetic field $\bf B_{ext}$. 
From the dependence of the spin polarization on $F_z$, motion direction, and external magnetic field $\bf B_{ext}$  we have determined  the \textit{IX} spin lifetime as well as the absolute magnitude and orientation of the SO fields in the DQW structures.
%paraRashba and Dresselhaus fields. %Furthermore,  the spin precession frequency in this case depends on the relative orientation between the external and SO field. 
In contrast to the spin patterns in collective regime addressed in Ref.~\cite{High_Nature_12}, we show that  transport and manipulation of \textit{IX}  ensembles reported here takes place in the drift-diffusive regime and can be potentially extended  down to  the single spin level.
%In contrast to the self-organized spin patterns resulting from collective effects,\cite{High_Nature_12} we show that the \textit{IX} spin transport and manipulation takes place in the drift-diffusive regime and can potentially applies to single spins.
% results were obtained in the of Ref.
%In addition to provide basic information about the \textit{IX} dynamics, these results demonstrate the feasibility of long-range transport and electrical control of \textit{IX} spins  in DQW structures.

In the following, we first describe the sample structure and the photoluminescence (PL) technique used to probe the transport of excitons spins. We then present experimental results demonstrating the transport and coherent precession of IX spins in the SO field  together with detailed studies of the spin dynamics as a function of propagation direction, bias, and  external magnetic fields. These results on spin transport  and SO effects are compared to the state-of-the art in the final sections of the manuscript.

% % ********************************************************************
% Figure1: FigIXSetup
% ********************************************************************

\section{Experimental Details}
\label{Experimental_Details}

The samples used in the experiment contain three sets of asymmetric GaAs DQWs grown by molecular beam epitaxy on a $n^{+}$-doped GaAs(001) substrate [cf.~Figs.~\ref{FigIXSetup}(a)-(b)]. Each DQW is formed by  a $d_{QW,1}=14$~nm (QW$_1$) and a $d_{QW,2}=17$~nm-thick QW (QW$_2$) separated by a thin (4~nm-thick) $Al_{0.3}Ga_{0.7}As$ barrier. Multiple DQWs were used in order to enhance the photon yield of the PL experiments.   The different QW widths enable the selective generation of direct excitons in either QW$_1$ or QW$_2$ by the appropriate selection of the excitation wavelength \cite{Stern_PRL101_257402_08}.   The electric field $F_{z}$ for the formation of \textit{IX}s was induced by a bias voltage $V_\mathrm{BIAS}$ applied  between the n-doped substrate and a thin (10 nm-thick) semitransparent Ti electrode (STE) deposited on the sample surface. 

The spectroscopic experiments were carried out  in a He bath cryostat at 2~K. Excitons were generated  using a right circularly polarized beam from a cw Ti:Sapphire laser resonant with the electron-heavy hole direct exciton (DX) transition of the wider (QW$_2$) QW. The laser was focused onto a spot with a diameter of 2.5~$\mu$m using a microscope objective. 
%Due to diffusion and repulsive interactions, the photo-excited \textit{IX}s cloud expands to distances considerably larger than the spot diameter.  
The PL around the laser spot was collected by the same objective, analyzed regarding the circular polarization, and then imaged on the entrance slit of a spectrometer connected to a cooled charge-coupled-device (CCD) camera. 
%In the PL imaging experiments using resonant excitation, the residual stray light from the laser was attenuated by a  long-pass filter introduced at the entrance of the spectrometer. 
The detected CCD images contain information about the energy (direction perpendicular to the slit) and spatial distribution of the PL intensity (along the slit). The spatial (coordinate $r$) dependence of the spin polarization was calculated according to $\rho_z(r) = \left[I_{\sigma^+}(r)-I_{\sigma^-}(r)\right]/\left[I_{\sigma^+}(r)+I_{\sigma^-}(r)\right]$, where $I_{\sigma^+}(r)$ ($I_{\sigma^-}(r)$)  denote the PL intensity at $r$ with right (left) circular polarization.

%%%%%%%%%%%%%%%%%%%%%%%%%%%%%%%%%%%%%%%%%%%%%%%%
\section{Results} 
\label{Results} 
 
\subsection{Coherent spin precession during transport}
\label{SpinTransport}
        
%\subsection{Bias dependence of the \textit{IX} lines}               

\begin{figure}[tbhp]
\begin{center}
\includegraphics[width=\linewidth, keepaspectratio=true,angle=0,clip]{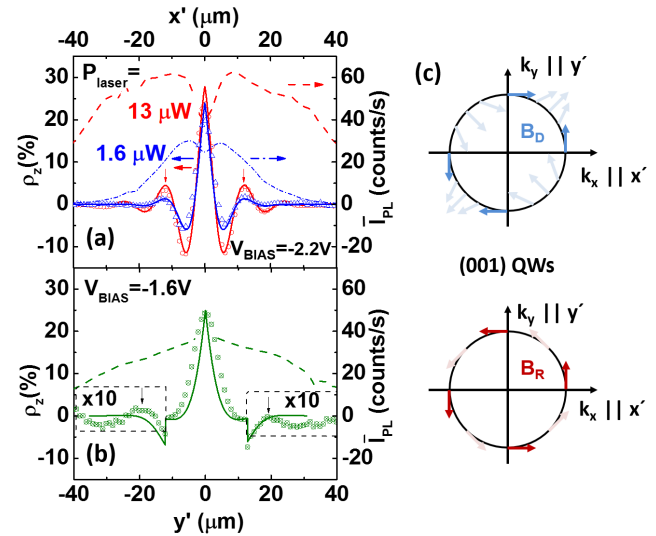}
\caption{(a) Profiles for the spin polarization ($\rho_s$, symbols) and for the average PL intensity ($\bar I_{PL}$, dashed and dot-dashed lines) for exciton transport along the $x'||[110]$ (symbols) recorded for laser excitation densities $P_\ell=1.6~\mu$W (open symbols) and $13~\mu$W (solid symbols). 
  and $[\bar110]$ directions (open symbols). 
(b) Corresponding profiles for transport along $y'||[\bar110]$ recorded for $P_\ell=6~\mu$W.  
The solid lines superimposed on the symbols are fits of the experimental results to Eq.~\ref{EqIX1}. 
(c) In-plane dependence of the Dresselhaus ($B_D$) and Rashba ($B_R$) contributions to the SO effective magnetic field $B_{SO}=B_D+B_R$. }
\label{FigIXSO}
\end{center}
\end{figure}

Figure~\ref{FigIXSetup}(c) compares PL images with right ($\sigma^+$) and left ($\sigma^-$) polarization recorded  with the spectrometer slit aligned with the sample crystallographic direction $x'=[110]$. The emission patterns at 1.515~eV and 1.525~eV arise, respectively, from excitons in the GaAs substrate and from the low energy tail of DXs in QW$_2$. In both cases, the emission is restricted to a region close to the excitation spot, thus indicating negligible diffusion distances for these excitations.  The line at 1.520~eV is associated with the emission of \textit{IX}s in the DQW structure, which in this case was subject to a bias $V_\mathrm{BIAS}=-2$~V. The emission extends to radii exceeding  40~$\mu$m - these long transport distances arise from the fact that the IXs are driven not only by  diffusion but also  by the drift forces induced by repulsive exciton-exciton dipolar interactions at the areas of high exciton density close to the excitation area. 
This assertion is supported by the strong dependence of the extension of the  \textit{IX} PL cloud  on the excitation density $P_l$, as displayed by the dashed and dash-dotted lines in Fig.~\ref{FigIXSO}(a). These curves displayed the spatial dependence of 
the total PL intensity (i.e.,  $[I_{\sigma^+}(r)+  I_{\sigma^-}(r)]$ determined from PL images for laser excitation powers of $P_l=13~\mu$W (as in Fig.~\ref{FigIXSetup}(c)) and 1.6~$\mu$W, respectively. The characteristic shape of the spatial PL profiles  are consistent with the observations of Ref.~\cite{Butov_N418_751_02}, which reported that the \textit{IX} PL reduces in the area of hot \textit{IX} close to the excitation spot. Furthermore, a closer look to  Fig.~\ref{FigIXSetup}(c) also reveals that the \textit{IX} emission energy red-shifts as one moves away from the excitation spot. In fact, the average \textit{IX} emission energy  at $x'=0~\mu$m is  $\sim0.5$~meV higher than  at $x'=75~\mu$m. This energetic blue-shift  is attributed to the strong repulsive  \textit{IX}-\textit{IX} interaction at the high \textit{IX} densities (estimated~\cite{RolandZimmermann_SSC144_395_07} to be of $2\times 10^{10}$~cm$^{-2}$)  close to $x'=0$.

The intensity of the circularly polarized \textit{IX} emission in Fig.~\ref{FigIXSetup}(c) is also spatially modulated with the maxima  for the $\sigma^+$ polarization corresponding to minima in the $\sigma^-$ polarization.
The symbols in Fig.~\ref{FigIXSO}(a) display the corresponding profiles for the spin  polarization $\rho_z$. 
%As will be further justified below, 
This spatial PL modulation  is attributed to the precession of the  \textit{IX} electron spins in the effective SO magnetic field $B_{SO}$ during their motion away from the excitation spot. 
% where the spins diffuse along $x'=[110]$. Results are shown  
%The dashed lines in  Fig.~\ref{FigIXSO}(a) display the spatial dependence of the average PL intensity (i.e.,  $(I_{\sigma^+}(r)+  I_{\sigma^-}(r))/2$ determined from the PL images (as in Fig.~\ref{FigIXSetup}(c)) recorded for two laser excitation intensities ($P_\ell=1.6$ and $13~\mu$W). 
%The profiles were recorded for IX propagating along the $x'=[110]$ direction. 
%The spatial extension of the emission region is considerably larger than the size of the excitation spot and increases with  $P_\ell$.  This behavior is attributed to the previously mentioned expansion due to repulsive exciton-exciton interactions.
%, which provides an extra driving force (in addition to normal diffusion) expelling the IXs from the high concentration region around the excitation spot.  
%
 The $\rho_z$ oscillations  extend over a range of radial distances shorter than the radius of the emission cloud. While the amplitude of the oscillations increases with $P_l$, their spatial period remains essentially constant. A similar  increase in spin polarization with excitation density was previously reported in Ref.~\cite{Larionov_PRB73_235329_06} and attributed to the onset of collective exciton effects. The fact that the spatial precession period does not change with the excitation density indicates, however, that the spin precession rates under the SO field does not depend on the \textit{IX} density. 
In addition, the larger $\rho_z$ amplitudes in our case is at least in part due to the shorter expansion times of the  \textit{IX} cloud resulting from \textit{IX}-\textit{IX} interactions under high $P_l$, which reduces the impact of spin scattering processes. 

The solid lines in Fig.~\ref{FigIXSO}(a) are  fits of the experimental results to damped oscillations with a spatial dependence given by:

\begin{equation}
\rho_s({\bf r}) = \rho_s(0) e^{-r/\ell_s} \cos{(2\pi r/\ell_{p}({\bf \hat r}))}.
\label{EqIX1}
\end{equation}

\noindent Here $\ell_{p}({\bf \hat r})$ is the spin spatial precession period for propagation along ${\bf \hat r}$, which is equal to the SO precession period $\ell_{so}$ in the absence of external magnetic fields.  $\ell_s$ denotes the effective spin transport decay length. Equation~(\ref{EqIX1}) describes  the  spin dynamics in a non-degenerate QW at low temperatures (i.e., for thermal energies less than the QW confinement energy), when 
%$k_BT<< \frac{\hbar^2}{2m_e } (\frac{\pi}{d_w})^2$, where $k_B$ is Boltzmann's constant, $m_e$ the effective electron mass, and $d_w$ the QW width), 
 spin precession rate $\Omega_{so}({\bf k}) = \frac{\mu_B B_{SO}}{\hbar}$ becomes proportional to the electron wave vector {\bf k}. Under this condition, the spin precession period (for ${\bf B_{ext}}=0$)   $\ell_{p} = \ell_{so} = \frac{m_e}{2\pi\hbar^2} \frac{k}{\Omega_{so}({\bf k})} $  only depends on direction ${\bf \hat r}$. The spin decay length $\ell_s$, in contrast, depends on temporal dynamics of the carrier motion as well as on the details of the spin scattering processes. The fits carried out with a single value for $\ell_p$ (i.e., independent of $r$) displayed in Eq.~\ref{EqIX1} reproduces very well the measured spin polarization profiles.

% ********************************************************************
% Figure 2:   FigIXSO: voltage dependence
% ********************************************************************

\subsection{Dependence on propagation direction}
\label{PropagationDependence}

%As mentioned in the introduction,
%Sec.~\ref{Introduction}, 
%the effective SO magnetic field $B_{SO}=B_D+B_R$ includes an  intrinsic (Dresselhaus, $B_D$) contribution associated with the lattice symmetry as well as  an electric field-dependent (Rashba, $B_R$) contribution  with  the momentum dependences illustrated in Fig.~\ref{FigIXSO}(c) for GaAs (001) QWs. 

The wave vector (${\bf k}$) dependence of the SO contributions $\bf B_D(k)$ and $\bf B_R(k)$ for a III-V QW is illustrated in Fig.~\ref{FigIXSO}(c). Here, we use a  reference frame defined by the axes  $x'||[110]$, $y'||[\bar110]$, and $z||[001]$. The associated  (temporal) electron spin precession frequencies $\bf \Omega_i$, $(i=D, R)$   can be written as a function of the wave vector ${\bf k}$ as:\cite{PVS272} 

\begin{equation}
\hbar{\bf \Omega_D}({\bf k})  = 
 \gamma \left( \frac{\pi}{d^\mathrm{eff}_{QW}} \right)^2
    \left[      \begin{array}{c}
     k_{y'}  \\
     k_{x'}  \\
     0
     \end{array}     \right] \quad\textit{and}
\label{EqIXA1}     
     \end{equation}
    
    \begin{equation}
     \quad
\hbar{\bf \Omega_R}({\bf k}) = 
2 F_z r_{41}
    \left[      \begin{array}{c}
         k_{y'} \\
        -k_{x'} \\
        0
     \end{array}    \right]    
\label{EqIXA2}              
\end{equation}

\noindent where $\gamma$ and $r_{41}$ are the Dresselhaus and Rashba spin constants, respectively. Equations~\ref{EqIXA1} and \ref{EqIXA2} apply for  small wave vectors (i.e., $|k_x'|,|k_x'|<<\frac{\pi}{d^\mathrm{eff}_{QW}}$. $d^\mathrm{eff}_\mathrm{QW}$ is the effective extension along $z$ of the electronic wave function  in the QW confining the electrons (QW$_2$ in the present case, as determined by the sign of V$_\mathrm{BIAS}$). $d^\mathrm{eff}_\mathrm{QW}$ is  determined by both the QW width and by the penetration depth  into the barrier layers.
 The corresponding amplitudes of the spatial precession frequencies  $\ell^{-1}_{so,i}$ can be derived from the precession distance required for a spin rotation by $2\pi$. The following expressions are obtained:\cite{Winkler03a}

\begin{equation}
\ell^{-1}_{so,D} =  \gamma \frac{m_e}{2\pi\hbar^3} \left( \frac{\pi}{d^\mathrm{eff}_\mathrm{QW}}\right)^2 \quad \textit{and}\quad   \ell^{-1}_{so,R} =\frac{m_e}{\pi\hbar^3}  r_{41}F_{z}.
\label{EqIX2}
\end{equation}

\noindent 
If ${\bf B_D}||{\bf B_R}$ the resulting spatial precession frequency will be given by $\ell^{-1}_{so} =  |\ell^{-1}_{so,D} \pm  \ell^{-1}_{so,R}|$, where the sign is defined by the relative orientation of the two fields. 

Equations~\ref{EqIXA1} and \ref{EqIX2} predicts that the spatial precession period $\ell_{so}$ depends both on transport direction and on bias electric field $F_z$. In order to check the in-plane anisotropy of $\bf B_{SO}$, we show in Fig.~\ref{FigIXSO}(b) spin polarization profiles recorded for \textit{IX} motion  along the $[1\bar10]$ surface direction (i.e., orthogonal to the one in Figs.~\ref{FigIXSetup}(c) and \ref{FigIXSO}(b)).  The period of the precession oscillations is much larger than for  the motion along the $[110]$ direction. The larger period, which in this case becomes comparable with the spin decay length $\ell_s$, makes the amplitude of the oscillations weaker than in Fig.~\ref{FigIXSO}(a).   The dependence of $\ell_{so}$ on transport direction is in agreement with the diagrams of Fig.~\ref{FigIXSO}(c):  while  ${\bf B_D}$ and ${\bf B_R}$ add for motion along $[110]$, they partially cancel each other for motion along $[\bar110]$, thus leading to a weaker SO field. %(and, therefore, to a lower spatial precession frequency). 
Interestingly, while a weaker $\bf B_{SO}$ reduces spin scattering, it does not necessarily improve the visibility of the  $\rho_z$ oscillations since one also has to satisfy the constraint $\ell_{so}<\ell_s$. On the other hand, a stronger $\bf B_{SO}$ reduces the spatial precession period. In this case, however, one needs to reduce the excitation spot to dimensions at least a factor of 2 smaller than the oscillation period in order to measure the oscillations. 

%As will be further discussed in Sec.~\ref{Discussions} the observation of clear precession oscillations in spatially resolved optical experiments results from the combination of the large SO field along $[110]$ (i.e., with a precession period $\ell_{so}<\ell_s$  comparable or less than the spin decay length) with the fast motion induced by the repulsive interaction forces.

%%, and the high spatial resolution of the optical experiments.

% ********************************************************************
% Figure 2:   FigIXBias: voltage dependence
\begin{figure}[tbhp]
\begin{center}
\includegraphics[width=\linewidth, keepaspectratio=true,angle=0,clip]{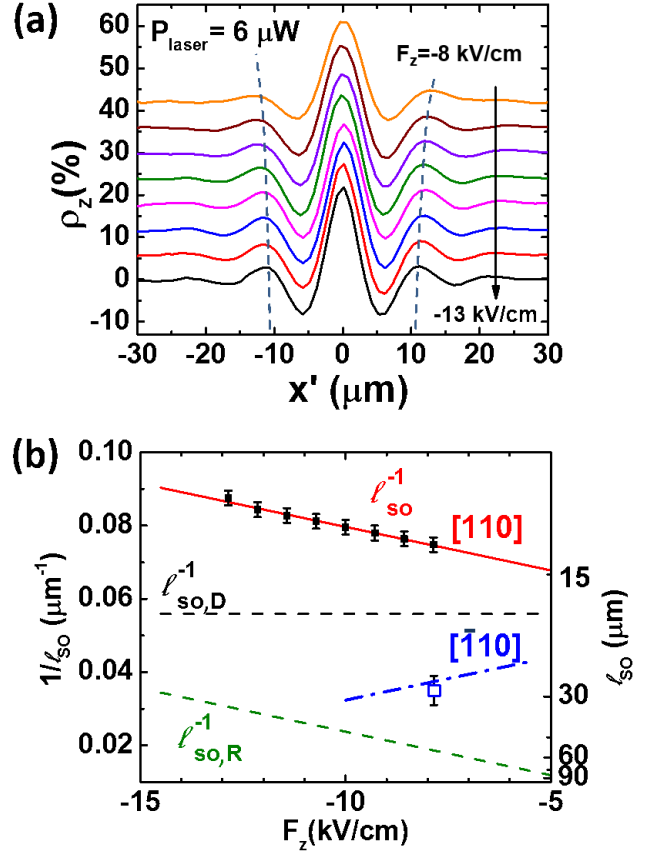}
\caption{(a) Profiles for the spin polarization ($\rho_s$, symbols) recorded for increasing electric fields $F_\mathrm{z}$  applied along the growth direction of the DQW structure. 
(b) Dependence of the spatial spin precession frequency $1/\ell_{so}$ on the electric field $F_{z}$ for transport along $x'|| [110]$ (solid symbols) and $y'||[\bar110]$ (open symbol). The solid line is a fit of the data along $[110]$ to $\ell^{-1}_{so} =  \ell^{-1}_{so,D} +  \ell^{-1}_{so,R}$ (cf.~Eq.~\ref{EqIX2}) with $\gamma=17.9$~eV\AA$^3$, $r_{41}=-8.5$~eV\AA$^2$, and $d^\mathrm{eff}_\mathrm{QW} = 21$~nm. The corresponding inverse spatial precession periods  $\ell^{-1}_{so,D}$ and $\ell^{-1}_{so,R}$ are illustrated by the dashed lines. The dot-dashed line shows the expected inverse precession period for motion along $[1\bar10]$. 
}
\label{FigIXBias}
\end{center}
\end{figure}
% ********************************************************************

\subsection{Electric spin control}
\label{BiasDependence}

%Further information about the SO interaction can be obtained from the dependence of $\rho_z$ on the electric field $F_z$,
% induced by the bias $V_\mathrm{BIAS}$ applied along the DQW growth direction, 
%which controls the Rashba component $B_R$. 

The electric control of the spin precession frequency is demonstrated by the series of profiles recorded for transport  along $[110]$  in Fig.~\ref{FigIXBias}(a). The amplitude $F_z$ of electric field was determined from the applied bias $V_\mathrm{BIAS}$ using the procedure described in Ref.~\cite{PVS260}. The spatial precession period $\ell_{so}$ reduces by over 15\% with increasing $F_{z}$, thus indicating an enhancement of the Rashba field $\bf B_R$ as well as of the total SO field $\bf B_{SO}$. 
%The data in Fig.~\ref{FigIXBias}(a) shows that the precession period can be electrically changed by over 15\%: 
Larger changes can be in principle be achieved by increasing the bias across the DQWs.  These, however, also reduce the PL yield and may lead to field-induced carrier leakage out of the DQW structure, thus hampering an efficient spin detection.\cite{PVS261}

The measured spatial precession frequencies ($\ell^{-1}_{so}$) are plotted as a function of the transverse electric field in Fig.~\ref{FigIXBias}(b).  The solid line is a fit of the data measured along $[110]$ to the expression: $\ell^{-1}_{so} =  \ell^{-1}_{so,D} +  \ell^{-1}_{so,R}$ (cf.~Eq.~\ref{EqIX2}), which gives $\gamma=17.9$~eV\AA$^3$, $r_{41}=-8.5$~eV\AA$^2$, and $d^\mathrm{eff}_\mathrm{QW} = 21$~nm (the signs for $\gamma$ and $r_{41}$ will be determined below. 
%in Sec.~\ref{MagneticDependence}).   
The corresponding inverse precession periods $\ell^{-1}_{so,D}$ and $\ell^{-1}_{so,R}$ for propagation along this direction are illustrated by the dashed lines.
For the applied bias, indirect excitons are created in the DQW structure  with the electrons confined within the wider (17~nm wide) QW. The value for $d^\mathrm{eff}_\mathrm{QW}$ used in the fits is compatible with an electronic wave function with a penetration depth of 2~nm in the barriers. As will be further discussed later,
% in Sec.~\ref{Discussions}, 
these SO parameters agree well with values reported in the literature.

The open symbol in Fig.~\ref{FigIXBias}{}(b) shows the inverse precession period measured for spin propagation along $y'$, where the Dresselhaus contribution changes in sign relative to the $x'$. The measured value agrees well with the calculated inverse periods $\ell^{-1}_{so}$ for this propagation along this direction (dash-dotted line).  Note that the change in relative signs of the Rashba and Dresselhaus effective magnetic fields can in principle also be attained for a fixed motion direction by simply reversing the bias polarity. For the asymmetric DQWs used in the present experiments, however,  a reversal in direction of $F_z$ moves the electrons to the narrow QW, thus also changing the magnitude of the Dresselhaus contribution. In addition, while the leakage current is very low for negative biases (a few nA), it increases considerably for high positive biases, thus indicating enhanced carrier leakage. 

%In the apresent sa As a consequence, the electrons and the holes will be confined in the opposite QWs with respect to the reverse bias condition, thus making difficult a straightforward comparison between the two cases.} The dot-dashed line displays the  $\ell^{-1}_{so}$ calculated for this direction using the fitting parameters listed above.

%The SO parameter $\gamma=17.9$~eV\AA$^3$  also compares well with those determined from electron spin transport  in $(100)$,\cite{PVS152}  $(110)$,\cite{PVS171} and $(111)$ QWs~\cite{PVS261} with comparable widths.  The Rashba parameter $r_{41}=8.5$~eV\AA$^2$, however, lies above the experimental values of  $4$ and $6\pm 1$~eV\AA$^2$ of Refs.~\cite{Eldridge_PRB82_45317_10} and \cite{PVS261}, respectively, as well as the calculated value of  $5.2\pm 1$~eV\AA$^2$ from Ref.~\cite{Winkler03a}.  While further investigations are required to clarify this difference, we speculate that it may be related to the asymmetric barriers in the DQW structure, which make the electronic wave function  more susceptible to applied electric fields.

\subsection{Magnetic field dependence}
\label{MagneticDependence}

An external magnetic field $\bf B_{\mathrm{ext}}$ applied along the DQW plane provides an additional degree of freedom to control the spin vector. The experiments were carried out under the configuration illustrated in Fig.~\ref{FigIXSetup}(a), where one maps the spins moving along $x'$ under an external field $\bf B_{\mathrm{ext}}||y'$ and collinear with $\bf B_{so}$.
Figure~\ref{FigIXB}(a) displays $\rho_z$ profiles along $x'$ recorded under different external magnetic fields. 
%The impact of $B_{\mathrm{ext}}$ on $\rho_z$ depends on the field strength. 
Large magnetic fields  ($|\bf B_{\mathrm{ext}}|>6$~mT) mainly induce a reduction of the overall spin polarization (Hanle effect),  from which one can extract the spin dephasing time $\tau_s$. For that purpose, we plot in Fig.~\ref{FigIXB}(b) the spin IX polarization at the excitation spot $\bar \rho_z$
%decay  integrated spin density $\bar\rho_s = \int_0^\infty |\rho_s(r)| dr$ 
as a function of $\bf B_{\mathrm{ext}}$.   
%$\bar\rho_s$ rather than the local $\rho_s$ has been used here in order to account for the inhomogeneous spin density resulting from the precession in the SO fields. 
The lines superimposed on the data points are fits to the Hanle expression:%~\cite{xx}

% ********************************************************************
% Figure 3:   FigIXB
\begin{figure}[tbhp]
\begin{center}
\includegraphics[width=\linewidth, keepaspectratio=true,angle=0,clip]{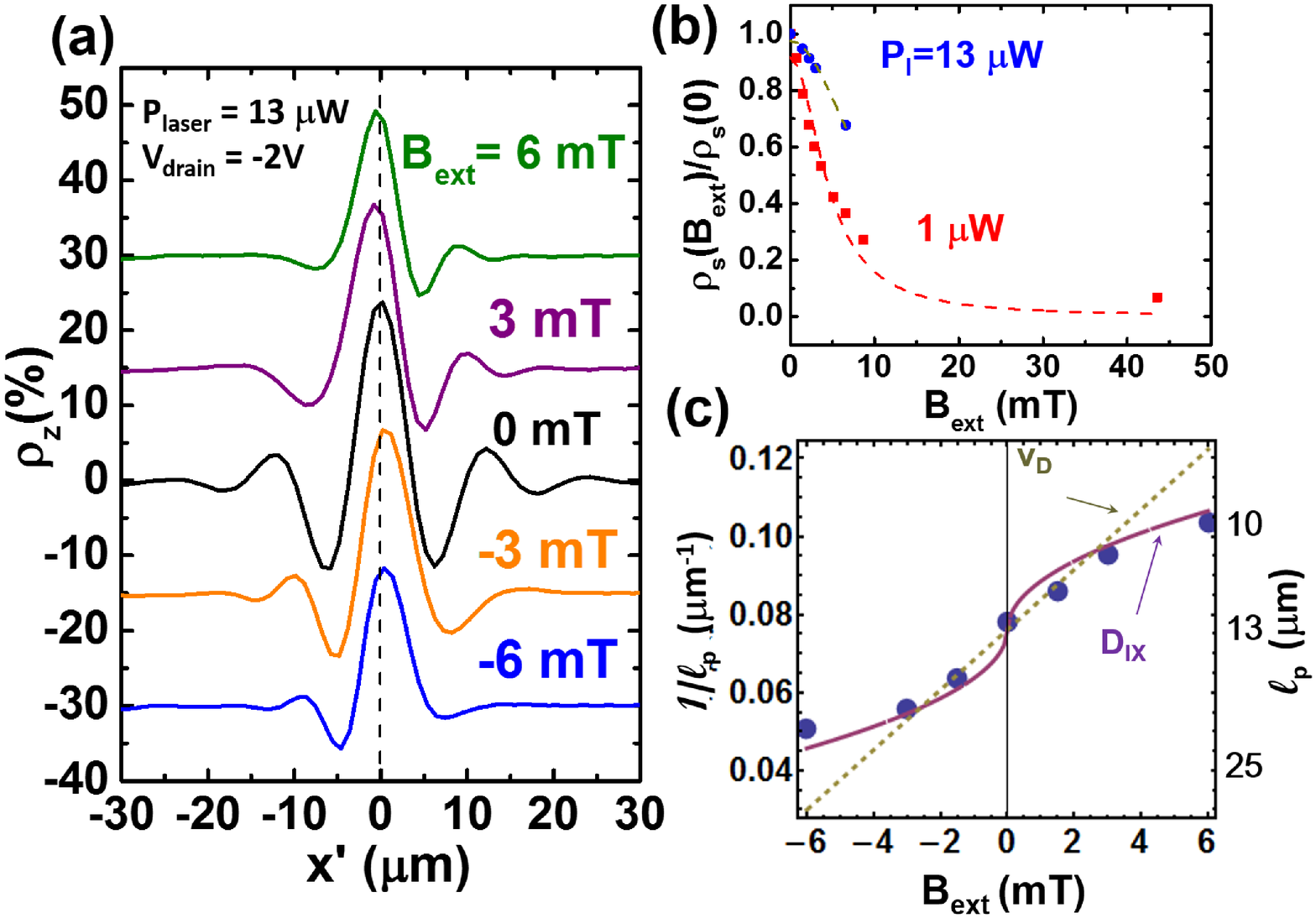}
\caption{(a) Spin polarization profiles as a function of an external magnetic field ($B_{\mathrm{ext}}$). The curves are stacked for clarity. 
%(b) Integrated spin polarization $\bar\rho_s = \int_0^\infty |\rho_s(r)| dr$ 
(b) $\rho_z$ close to the excitation spot as a function of $B_{\mathrm{ext}}$ measured for optical excitation densities $P_l=13~\mu$W (squares) and $1~\mu$W (dots).  The dashed lines display fits to Eq.~\ref{EqIXHanle}, from which one extracts spin lifetimes of 6 and 3~ns, respectively. (c) Dependence of the inverse precession period ($1/\ell_{p} = 1/\ell_{so}+1/\ell_{B}$, where $\ell_{so}$ and $\ell_{B}$ are the precession period around the SO and external fields, respectively) on $B_{\mathrm{ext}}$ (symbols). The lines are the predictions from the diffusion ($D_{IX}$) and drift ($v_D$) models discussed in the text.}
\label{FigIXB}
\end{center}
\end{figure}
% ********************************************************************

% --------------------------------------------------------------------------------------
\begin{equation}
\bar\rho_z(B_{\mathrm{ext}}) = \frac{\bar \rho_z(0) }{ 1 + (\omega_L\tau*)^2} \quad\textit{with}\quad \hbar\omega_L=g_e\mu_B B_{\mathrm{ext}}.
\label{EqIXHanle}
\end{equation}
% --------------------------------------------------------------------------------------
\noindent where $\bar\rho_z$ is the degree of polarization at the generation point, $\frac{1}{\tau*}=\frac{1}{\tau_{IX}}+\frac{1}{\tau_s}$ with $\tau_{IX}$ the indirect exciton lifetime.
By using the conduction band g-factor $|g_e|=0.44$ and considering $\tau_{IX} >> \tau_{s}$ , one extracts from the fits spin lifetimes of 3 and 6~ns for excitation powers of $P_l=1~\mu$W and $13~\mu$W, respectively. These lifetimes are considerably longer than the typical carrier scattering times, thus indicating that the spin dynamics observed here takes place in diffusive (or strong scattering) regime.
 
Weak magnetic fields, in contrast, mainly affect the spin spatial precession frequency, as indicated by the symbols in Fig.~\ref{FigIXB}(c). In contrast to the SO field, which reverses with carrier momentum, the external field has a fixed direction and its action on the spin vector only depends on time. These features can be exploited to determine the absolute orientation of $\bf B_{so}$.
%, as illustrated in Fig.~\ref{FigIXB}(a). 
When $\bf B_{\mathrm{ext}}$ is oriented along $+y'$ (upper curves for $\bf B_{\mathrm{ext}} > 0$ in Fig.~\ref{FigIXB}(a)), the spatial spin precession rate  increases (reduces) for spins moving along $+x'$ ($-x'$). The opposite behavior is observed when the orientation of  $\bf B_{ext}$ reverses (curves for $\bf B_{\mathrm{ext}}<0$). From these results, together with  
%the observation that the spatial precession frequency along the $x'$ increases increasing the electric field (see Sec. 3.3), 
 the increase of the spatial precession frequency along the $x'$ with $|F_z|$ (cf. Fig.~\ref{FigIXBias}(a), 
we unambiguously establish that both the Rashba and Dresselhaus spin orbit fields point towards $y'$ when spins move along $x'$.  The Dresselhaus splitting constant $\gamma$ in Eq.~\ref{EqIXA1}  is thus positive while the Rashba parameter  $r_{14}$ is negative.

%The precession of moving spins around $B_{so}$ and $B_{\mathrm{ext}}$ depends only on the propagation distance and time, respectively. 
The different natures of the two fields allows us to extract information about the precession dynamics during motion and their impact on the precession period $\ell_p$. Here, we first note that the different $\ell_p$ values in Fig.~\ref{FigIXB}(a)  along $+x'$ and $-x'$ implies that both fields (i.e., $\bf B_{so}$ and $\bf B_{\mathrm{ext}}$) must act simultaneously on the spins during the motion. This becomes clear if one considers the hypothetical situation, where the spins initially move to the final position $x'$ within a very short time (thereby precessing only around $\bf B_{so}({\bf k})$) and then remain trapped there until recombination (thus precessing only around $\bf B_{\mathrm{ext}}$). 
%In this case, the precession period would remain the same for spin motion along  both the  positive and negative $x'$ axes, since $B_{\mathrm{ext}}$ would just add a fixed (i.e., position independent) phase to the spin precession angle. 
In this case, $\bf B_{\mathrm{ext}}$ would simply add a fixed (i.e., position independent) phase to the spatial oscillations determined by $\bf B_\mathrm{so}$ and $\ell_p$  would be the same for spin motion along  both the  positive and negative $x'$ axes, in contrast to the results displayed in Fig.~\ref{FigIXB}(a).  

Since $\bf B_{\mathrm{ext}}$ and $\bf B_\mathrm{so}$ are collinear in Fig.~\ref{FigIXB}(a), the inverse precession period can be  written as
$\ell^{-1}_p = \ell^{-1}_{so} + \ell^{-1}_B(\bf B_{\mathrm{ext}})$, where $\ell^{-1}_B(\bf B_{\mathrm{ext}})$ is the contribution associated with the external field. In order to account for the dependence of $\ell^{-1}_p$  
 on $\bf B_{\mathrm{ext}}$ one needs information about the spin position $x'$ as a function of time ($t$). We consider  the following two extreme situations. First, if the motion is mainly driven by the IX-IX repulsion forces at the excitation region, the carriers are expected to move with an approximately constant velocity $v_D$, yielding $\ell^{-1}_B( B_{\mathrm{ext}})=\ell^{-1}_{B,v_D} = \frac{g \mu_B}{2 \pi \hbar} \frac{\bf B_{\mathrm{ext}} }{ {v_D}  }$. This model predicts the linear dependence of $\ell^{-1}_p$ on field  indicated by the  dashed line in Fig.~\ref{FigIXB}(c). The very small values used for  $v_D=80$~m/s, which corresponds to only ~$\sim$1\% of the average exciton thermal velocity at 2~K, probably arise from the fact that diffusion has been completely neglected. While the behavior for small fields is well accounted for, the model fails to reproduce the deviations from linearity observed as the $|B_{\mathrm{ext}}|$ increases. The second extreme situation occurs if motion in purely driven by diffusion, in which case  $x'=\sqrt{D_{IX}t}$, where $D_{IX}$ is the IX diffusion coefficient. A straight-forward calculation yields $\ell^{-1}_B(B_{\mathrm{ext}})=\ell^{-1}_{B,D} = \sqrt{\frac{g \mu_B}{2 \pi \hbar} \frac{\bf B_{\mathrm{ext}} }{D_{IX} }}$, leading to the behavior depicted by the solid line in Fig.~\ref{FigIXB}(c). The calculations were carried out using an exciton diffusion coefficient $D_{IX}=40\times10^{-4}$~cm$^2$/s, which is approximately a factor of 2 larger than the one expected from Ref.~\cite{Voros_PRL94_226401_05} for diffusion in a 15~nm wide QW. Despite its simplicity, this approximation reproduces reasonably well the measured behavior for large field amplitudes.
%over the whole range of investigated fields. 
A good fit requires, however,  values for $D_{IX}$ higher that those expected from Ref.~\cite{Voros_PRL94_226401_05}. This becomes necessary to  reduce the IX transport (and precession) time and,  in that way, compensate for the errors arising neglecting of drift effects. A more precise description of the spin dynamics by taking into account drift and diffusion is presently under study.

%\begin{eqnarray}
%\ell^{-1}_{B,D}    &=& \sqrt{ \frac{ g \mu_B B_{\mathrm{ext}} }{2 \pi \hbar D  }} \\ 
%\ell^{-1}_{B,v_D} &=& \frac{g \mu_B B_{\mathrm{ext}} }{2 \pi \hbar {v_D}  }
%\label{EqIXlB}
%\end{eqnarray}

% -------------------------------------------------------------------------------------------------------
\section{Discussions}
\label{Discussions}

The experimental results of the previous section unambiguously show the coherent precession of optically excited spins in the effective magnetic field $\bf B_{so}$ induced by the SO interaction. The spin precession period depends on the in-plane motion direction - it is essentially independent of the \textit{IX} density and can be controlled by external electric and  magnetic fields.   The measured spin lifetimes (of a few ns) indicate that transport  takes place in the drift-diffusive regime. The spin dynamics is well-described by taking into account the SO parameters as well as the g-factor for electrons, thus showing that it is determined by the \textit{IX} electron spins.
In fact, the SO parameter $\gamma=17.9$~eV\AA$^3$   compares well with those determined from electron spin transport  in $(100)$,\cite{PVS152}  $(110)$,\cite{PVS171} and $(111)$ QWs~\cite{PVS261} with comparable widths.  The magnitude of the Rashba parameter $r_{41}=-8.5$~eV\AA$^2$, however, lies above the experimental values of  $4$ and $6\pm 1$~eV\AA$^2$ of Refs.~\cite{Eldridge_PRB82_45317_10} and \cite{PVS261}, respectively, as well as the calculated value of  $5.2\pm 1$~eV\AA$^2$ from Ref.~\cite{Winkler03a}.  While further investigations are required to clarify this difference, we speculate that it may be related to the asymmetric barriers in the DQW structure, which make the electronic wave function  more susceptible to the applied electric fields.

 The coherent transport and precession of electron spins in the SO field in the drift-diffusive transport  regime has been previously reported for  bulk GaAs layers~\cite{Kato03a,Beck05a} as well as  QW structures.~\cite{Kato03a,PVS152,PVS171,PVS240,PVS265}  
%structures.~\cite{Kato03a,PVS152,PVS171,PVS240,PVS265,Meier_NatPhys3_650_07}  
In these experiments, electron transport has been driven by an external electric (or piezoelectric field): in this way, the spins could be moved over microscopic distances before dephasing (assumed to be due to the D'yakonov-Perel' mechanism~\cite{Dyakonov_SPSS13_3023_72}) sets in.  To our knowledge, however, the coherent  precession of spins injected in an undoped QWs in  the absence of an electric driving field demonstrated here has so far not been reported. 

The observation of coherent spin precession in the SO field requires small ratios $\ell_{so}/\ell_{s}$ (cf.~Eq.~\ref{EqIX1}). Here, $\ell_{so} \sim a_{so}\langle \Omega_{so}\rangle$ ($a_{so}$ is a proportionality constant and the $\langle\rangle$'s denote thermal averages over the  transport path) only depends on the strength of the SO interaction.  The spin decay length  $\l_{s}$,  in contrast, depends on transport time. For purely diffusive motion in the D'yakonov-Perel' regime, 
$\l_{s}\sim \frac{D_{IX}}{2r} \frac{1}{ \tau_c \langle {\Omega^2_{so}}\rangle}$, where $\tau_c$ is the carrier scattering time and $D_{IX}$ the diffusion coefficient,   reduces with transport distance $r$. As a result, the requirement 
$\ell_{so}/\ell_{s} \sim \frac{2  \tau_c}{D_{IX}}
\frac{\langle {\Omega_{so}}\rangle}{\langle {\Omega^2_{so}}\rangle} a_{so} r< 1$  becomes increasingly difficult to be satisfied for long transport distances $r$ (as well as for strong SO fields). In the present experiments, the expansion forces due to \textit{IX}-\textit{IX} repulsive interactions appear to have a fundamental role, since they considerably reduce the transport time, thus increasing $\ell_s$ and enabling the observation of precession oscillations.

Coherent  spin patterns resulting from precession in the SO field have been reported for macroscopically ordered \textit{IX}  phases.\cite{High_PRL110_246403_13,High_Nature_12}  
%These authors observed spin texture with a four-fold angular symmetry, which were attributed to the precession of exciton spins under the Dresselhaus SO field.  
%These experiments have been carried out at temperature below 2~K,\cite{High_arXiv_11} where IX's are expected to form a macroscopic ordered state.\cite{High_Nature_12} 
The long spin coherence lengths required for their formation have been attributed to the reduced scattering in the ordered state, which enables the exciton transport in the ballistic regime.  The results presented in the previous section show that  long spin lifetimes and transport lengths,  as well as coherent precession in the SO field can also be achieved in the drift-diffusive regime. 
This behavior is followed occurs over  a wide range of exciton densities, thus indicating that it is not associated with collective effects. Interestingly,  the spatial extent of the precession oscillations reported here is comparable to the ones for the self-organized spin patterns.\cite{High_PRL110_246403_13}    
There are additional fundamental differences between the two cases. While the spin pattern observed in Ref.~\cite{High_PRL110_246403_13} results from self-organization and does not rely on the initial exciton polarization, the ones presented here appear as a consequence of the coherent transport of optically injected spins. In addition, the dynamics observed here is associated with electron spin precession, while Ref.~\cite{High_PRL110_246403_13}  has invoked the simultaneous evolution of hole and electron spins.

Finally, as in most of the reports of coherent  precession of optically injected spins in the SO field, the absence of hole spin contributions to the spin dynamics is attributed to the short hole spin coherence times as well as the spatial separation of the electron and hole wave function in the DQW structure.
%One interesting question regards the  that may appear in the present case, where we have 
Since we have excited and probed \textit{IX} spins, an interesting question is whether the electrons and hole remain bound during transport. Although we cannot present a conclusive evidence for this behavior, we mention that the \textit{IX} spins have been excited with small excess energy (approx. 10~meV), thus reducing the probability of exciton dissociation followed by the spatial separation of electrons and holes. In addition, the \textit{IX} spatial  profiles (cf., e.g.,~ Fig.~\ref{FigIXSetup}(c)) do not show the characteristic rings observed when electrons and holes diffuse independently.

\section{Conclusions}
\label{Conclusions}

We have presented experimental evidence for the coherent precession of exciton spins during transport in the effective magnetic field induced by the SO interaction. The precession dynamics is essentially independent of the IX density, but can be controlled by an external bias (via changes in the SO field) or magnetic field. The spin dynamics are well explained by taking into account electron spin precession, with no noticeable hole spin contribution.  The technique used by us allows for the control and discrimination of precession effects arising from the Rashba and Dresselhaus SO contributions, as well as  those induced by an external magnetic field. In this way, we were able to directly determine the absolute value and sign of the spin splitting parameters for the   Rashba and Dresselhaus SO mechanisms.  In addition,  the spin dynamics in the presence of an external magnetic field yield information about the precession dynamics during transport. The long spin transport distances as well as the control of the spin vector open the way for the use of IX spins to encode and manipulate quantum information in opto-electronic semiconductor devices. 

\section{Acknowledgements}
\label{Acknowledgements}
We thank Manfred Ramsteiner for discussions and comments on the manuscript. We are indebted to S. Rauwerdink, W. Seidel, S. Krau\ss, and A. Tahraoui for the assistance in the preparation 
of the samples. We also acknowledge financial support from the German DFG (grant 
SA-598/9).

\vspace{ 1 cm}

%\vskip 1 cm
%\noindent{\bf References}

\def\litdir{x:/sawoptik_databases/jabref}
%\def\litdir{y:/myfiles/jabref}
%\bibliography{\litdir/literature,\litdir/mypapers}

\begin{thebibliography}{99}

\bibitem{Sivalertporn_PRB85_45207_12}
K.~Sivalertporn, L.~Mouchliadis, A.~L. Ivanov, R.~Philp, and E.~A. Muljarov.
\newblock Direct and indirect excitons in semiconductor coupled quantum wells
  in an applied electric field.
\newblock {\em Phys. Rev. B}, 85:045207, Jan 2012.

\bibitem{Schinner_PRL110_127403_13}
G.~J. Schinner, J.~Repp, E.~Schubert, A.~K. Rai, D.~Reuter, A.~D. Wieck, A.~O.
  Govorov, A.~W. Holleitner, and J.~P. Kotthaus.
\newblock Confinement and interaction of single indirect excitons in a
  voltage-controlled trap formed inside double {InGaAs} quantum wells.
\newblock {\em Phys. Rev. Lett.}, 110:127403, Mar 2013.

\bibitem{PVS239}
Kobi Cohen, Ronen Rapaport, and Paulo~V. Santos.
\newblock Remote dipolar interactions for objective density calibration and
  flow control of excitonic fluids.
\newblock {\em Phys. Rev. Lett.}, 106(12):126402, Mar 2011.

\bibitem{Butov_N418_751_02}
L.~V. Butov, A.~C. Gossard, and D.~S. Chemla.
\newblock Macroscopically ordered state in an exciton system.
\newblock {\em Nature}, 418:751, 2002.

\bibitem{High_Nature_12}
A.~A. High, J.~R. Leonard, A.~T. Hammack, M.~M. Fogler, L.~V. Butov, A.~V.
  Kavokin, K.~L. Campman, and A.~C. Gossard.
\newblock Spontaneous coherence in a cold exciton gas.
\newblock {\em Nature}, 483:584, March 2012.

\bibitem{Shilo_NC4_2335_13}
Yehiel Shilo, Kobi Cohen, Boris Laikhtman, Ken West, Loren Pfeiffer, and Ronen
  Rapaport.
\newblock Particle correlations and evidence for dark state condensation in a
  cold dipolar exciton fluid.
\newblock {\em Nature Communications}, 4:2335, 2013.

\bibitem{High_S321_229_08}
Alex~A. High, Ekaterina~E. Novitskaya, Leonid~V. Butov, Micah Hanson, and
  Arthur~C. Gossard.
\newblock Control of exciton fluxes in an excitonic integrated circuit.
\newblock {\em Science}, 321:229, 2008.

\bibitem{Winbow_NL7_1349_07}
Alexander~G. Winbow, Aaron~T. Hammack, Leonid~V. Butov, and Arthur~C. Gossard.
\newblock Photon storage with nanosecond switching in coupled quantum well
  nanostructures.
\newblock {\em Nano Lett.}, 7(5):1349--1351, 2007.

\bibitem{Grosso_NP3_577_09}
G.~Grosso, J.~Graves, A.~T. Hammack, A.~High, L.~V. Butov, M.~Hanson, and A.~C.
  Gossard.
\newblock Excitonic switches operating at around 100 k.
\newblock {\em Nat. Photonics}, 3(10):577--580, October 2009.

\bibitem{Voros_PRL94_226401_05}
Z.~V{\"o}r{\"o}s, R.~Balili, D.W. Snoke, L.~Pfeiffer, and K.~West.
\newblock Long-distance diffusion of excitons in double quantumwell structures.
\newblock {\em Phys. Rev. Lett.}, 94:226401, 2005.

\bibitem{Gaertner_APL89_052108_06}
A.~G{\"a}rtner, A.~W. Holleitner, J.~P. Kotthaus, and D.~Schuh.
\newblock Drift mobility of long-living excitons in coupled {GaAs} quantum
  wells.
\newblock {\em Appl. Phys. Lett.}, 89:052108, 2006.

\bibitem{Winbow_PRL106_196806_11}
A.~G. Winbow, J.~R. Leonard, M.~Remeika, Y.~Y. Kuznetsova, A.~A. High, A.~T.
  Hammack, L.~V. Butov, J.~Wilkes, A.~A. Guenther, A.~L. Ivanov, M.~Hanson, and
  A.~C. Gossard.
\newblock Electrostatic conveyer for excitons.
\newblock {\em Phys. Rev. Lett}, 106:196806, 2011.

\bibitem{PVS177}
J.~Rudolph, R.~Hey, and P.~V. Santos.
\newblock Long-range exciton transport by dynamic strain fields in a {GaAs}
  quantum well.
\newblock {\em Phys. Rev. Lett.}, 99:047602, 2007.

\bibitem{PVS260}
S.~S.~Lazi\'{c}, A.~Violante, K.~Cohen, R.~Hey, R.~Rapaport, and P.~V. Santos.
\newblock Scalable interconnections for remote indirect exciton systems based
  on acoustic transport.
\newblock {\em Phys. Rev. B}, 89:085313, Feb 2014.

\bibitem{PVS266}
A.~Violante, K.~Cohen, S.~Lazi\'c, R.~Hey, R.~Rapaport, and P.~V. Santos.
\newblock Indirect exciton dynamics in moving potentials.
\newblock {\em New J. Phys.}, 89:085313, 2014.

\bibitem{Zutic_ROMP76_323_04}
I.~Zutic, J.~Fabian, and S.~D. Sarma.
\newblock Spintronics: Fundamentals and applications.
\newblock {\em Rev. Mod. Phys.}, 76:323, 2004.

\bibitem{Bir_SovPhysJETP42_705_75}
A.G.~Aronov G.L.~Bir and G.E. Pikus.
\newblock {\em Sov. Phys. JETP}, 42:705, 1975.

\bibitem{Bir74a}
G.~L. Bir and G.~E. Pikus.
\newblock {\em Symmetry and Strain-Induced Effects in Semiconductors}.
\newblock John Wiley \& Sons, New York, 1974.

\bibitem{Larionov_JL71_117_00}
AV~Larionov, VB~Timofeev, J~Hvam, and C~Soerensen.
\newblock Collective behavior of interwell excitons in gaas/algaas double
  quantum wells.
\newblock {\em JETP Lett.}, 71(3):117--122, 2000.

\bibitem{Larionov_PRB73_235329_06}
A.~V. Larionov, V.~E. Bisti, M.~Bayer, J.~Hvam, and K.~Soerensen.
\newblock Coherent spin dynamics of an interwell excitons gas in gaas/algaas
  coupled quantum wells.
\newblock {\em Phys. Rev. B}, 73:235329, 2006.

\bibitem{Kowalik-Seidl_APL97_11104_10}
K.~Kowalik-Seidl, X.~P. V\"{o}gele, B.~N. Rimpfl, S.~Manus, J.~P. Kotthaus,
  D.~Schuh, W.~Wegscheider, and A.~W. Holleitner.
\newblock Long exciton spin relaxation in coupled quantum wells.
\newblock {\em Appl. Phys. Lett.}, 97(1):011104, 2010.

\bibitem{Dyakonov_SPJETP_33_1053}
M.~I. D'yakonov and V.~I. Perel'.
\newblock Spin orientation of electrons associated with the interband
  absorption of light in semiconductors.
\newblock {\em SPJETP}, 33:1053--1059, 1971.

\bibitem{Dyakonov_Springer-Verlag_08}
Mikhail~I. Dyakonov.
\newblock {\em Spin physics in semiconductors}.
\newblock Springer, 2008.

\bibitem{Dresselhaus55a}
G.~Dresselhaus.
\newblock Spin-orbit coupling effects in zinc blende structures.
\newblock {\em Phys. Rev.}, 100:580--586, Oct 1955.

\bibitem{Rashba_SPSS2_1109_60}
E.~I. Rashba.
\newblock {\em Sov. Phys. Solid State}, 2:1109, 1960.

\bibitem{Poggio_PhysRevB70_121305_04}
M.~Poggio, G.~M. Steeves, R.~C. Myers, N.~P. Stern, A.~C. Gossard, and D.~D.
  Awschalom.
\newblock Spin transfer and coherence in coupled quantum wells.
\newblock {\em Phys. Rev. B}, 70:121305(R), 2004.

\bibitem{Larionov_PRB78_33302_08}
A.V. Larionov and L.E. Golub.
\newblock Electric field control of spin-orbit splittings in gaas/algaas
  coupled quantum wells.
\newblock {\em Phys. Rev. B}, 78:033302, 2008.

\bibitem{Burt_PRL79_337_97}
E.~A. Burt, R.~W. Ghrist, C.~J. Myatt, M.~J. Holland, E.~A. Cornell, and C.~E.
  Wieman.
\newblock Coherence, correlations, and collisions: What one learns about
  bose-einstein condensates from their decay.
\newblock {\em Phys. Rev. Lett.}, 79:337--340, 1997.

\bibitem{Timofeev_PhysRevB60_8897_99}
V.~B. Timofeev, A.~V. Larionov, M.~Grassi Alessi, M.~Capizzi, A.~Frova, and
  J.~M. Hvam.
\newblock Charged excitonic complexes in gaas/al0.35ga0.65as p-i-n double
  quantum wells.
\newblock {\em Phys. Rev. B}, 60:8897--8901, 1999.

\bibitem{Leonard_NanoLett9_4204_09}
J.~R. Leonard, Y.~Y. Kuznetsova, Sen Yang, L.~V. Butov, T.~Ostatnick,
  A.~Kavokin, and A.~C. Gossard.
\newblock Spin transport of excitons.
\newblock {\em Nano Lett.}, 9:4204--4208, 2009.

\bibitem{High_PRL110_246403_13}
A.~A. High, A.~T. Hammack, J.~R. Leonard, Sen Yang, L.~V. Butov,
  T.~Ostatnick\'y, M.~Vladimirova, A.~V. Kavokin, T.~C.~H. Liew, K.~L. Campman,
  and A.~C. Gossard.
\newblock Spin currents in a coherent exciton gas.
\newblock {\em Phys. Rev. Lett.}, 110:246403, Jun 2013.

\bibitem{Kavokin_PRB88_195309_13}
A.~V. Kavokin, M.~Vladimirova, B.~Jouault, T.~C.~H. Liew, J.~R. Leonard, and
  L.~V. Butov.
\newblock Ballistic spin transport in exciton gases.
\newblock {\em Phys. Rev. B}, 88:195309, Nov 2013.

\bibitem{Stern_PRL101_257402_08}
M.~Stern, V.~Garmider, E.~Segre, M.~Rappaport, V.~Umansky, Y.~Levinson, and
  I.~Bar-Joseph.
\newblock Photoluminescence ring formation in coupled quantum wells: Excitonic
  versus ambipolar diffusion.
\newblock {\em Phys. Rev. Lett.}, 101(25):257402, 2008.

\bibitem{RolandZimmermann_SSC144_395_07}
Christoph~Schindler Roland~Zimmermann.
\newblock Exciton–exciton interaction in coupled quantum wells.
\newblock {\em Solid State Commun.}, 144:395, 2007.

\bibitem{PVS272}
A.~Hern{\'a}ndez-M{\'i}nguez, K.~Biermann, R.~Hey, and P.~V. Santos.
\newblock Spin transport and spin manipulation in (110) and (111) {GaAs}
  quantum wells.
\newblock {\em Phys. Status Solidi B}, in print:--, 2013.

\bibitem{Winkler03a}
R.~Winkler.
\newblock {\em Spin-Orbit Coupling Effects in Two-Dimensional Electron and Hole
  Systems}, volume 191.
\newblock Springer, Berlin, 2003.

\bibitem{PVS261}
A.~Hern{\'a}ndez-M{\'i}nguez, K.~Biermann, R.~Hey, and P.V. Santos.
\newblock Electrical suppression of spin relaxation in {GaAs(111)B} quantum
  wells.
\newblock {\em Phys. Rev. Lett.}, 109:266602, 2012.

\bibitem{PVS152}
James A.~H. Stotz, Rudolf Hey, Paulo~V. Santos, and K.~H. Ploog.
\newblock Coherent spin transport via dynamic quantum dots.
\newblock {\em Nat. Mater.}, 4:585, 2005.

\bibitem{PVS171}
O.~D.~D. {Couto, Jr.}, F.~Iikawa, J.~Rudolph, R.~Hey, and P.~V. Santos.
\newblock Anisotropic spin transport in (110) {GaAs} quantum wells.
\newblock {\em Phys. Rev. Lett.}, 98:036603, 2007.

\bibitem{Eldridge_PRB82_45317_10}
P.~S. Eldridge, W.~J.~H. Leyland, P.~G. Lagoudakis, R.~T. Harley, R.~T.
  Phillips, R.~Winkler, M.~Henini, and D.~Taylor.
\newblock Rashba spin-splitting of electrons in asymmetric quantum wells.
\newblock {\em Phys. Rev. B}, 82(4):045317, Jul 2010.

\bibitem{Kato03a}
Y.~Kato, R.~C. Myers, A.~C. Gossard, and D.~D. Awschalom.
\newblock Coherent spin manipulation without magnetic fields in strained
  semiconductors.
\newblock {\em Nature}, 427:50, 2004.

\bibitem{Beck05a}
M.~Beck, C.~Metzner, S.~Malzer, and G.~H. Doehler.
\newblock Spin lifetimes and strain-controlled spin precession of drifting
  electrons in zinc blende type semiconductors.
\newblock {\em Europhys. Lett.}, 75:597, 2006.

\bibitem{PVS240}
H.~Sanada, T.~Sogawa, H.~Gotoh, K.~Onomitsu, M.~Kohda, J.~Nitta, and P.~V.
  Santos.
\newblock Acoustically induced spin-orbit interactions revealed by
  two-dimensional imaging of spin transport in {GaAs}.
\newblock {\em Phys. Rev. Lett.}, 106(21):216602, May 2011.

\bibitem{PVS265}
H.~Sanada, Y.~Kunihashi, H.~Gotoh, K.~Onomitsu, M.~Kohda, J.~Nitta, P.~V.
  Santos, and T.~Sogawa.
\newblock Manipulation of mobile spin coherence using magnetic-field-free
  electron spin resonance.
\newblock {\em Nat. Phys.}, 9:280, March 2013.

\bibitem{Dyakonov_SPSS13_3023_72}
M.~I. D'yakonov and V.~I. Perel'.
\newblock Spin relaxation of conduction electrons in noncentrosymmetric
  semiconductors.
\newblock {\em Sov. Phys. Solid State}, 13:3023--3026, 1972.

\end{thebibliography}
%\end{document}

%%%%%%%%%%%%%%%%%%%%%%%%%%%%%%%%%%%%%%%%%%%%%%%%%

%%%%%%%%%%%%%%%%%%%%%%%%%%%%%%%%%%%%%%%%%%%%%%%%%

\end{document}